# Sub-kelvin Andreev reflection spectroscopy of superconducting gaps in FeSe

D. L. Bashlakov[1], N. V. Gamayunova[1], L. V. Tyutrina[1], J. Kačmarčik[2], P. Szabó[2], P. Samuely[2]
Yu. G. Naidyuk[1]

[1]*B. Verkin Institute for Low Temperature Physics and Engineering, NAS of Ukraine, 61103 Kharkiv, Ukraine*

[2]*Centre of Low Temperature Physics, Institute of Experimental Physics, Slovak Academy of Sciences, Watsonova 47, SK-04001 Košice, Slovakia*

## Abstract

Point contact Andreev reflection studies have been conducted on FeSe single crystals by lowering the temperatures down to 0.5 K. The point contact Andreev reflection spectra were analyzed in the framework of the two-band model. As a result, the presence of two anisotropic superconducting gaps in FeSe were certainly established and their BCS-like temperature dependencies were obtained. The weights of each gap have been determined and the anisotropy parameter has been calculated. It is shown, that sub-kelvin temperatures are necessary to ascertain details of the superconducting gap structure, especially for multiband materials when Andreev reflection spectroscopy is applied.

## Introduction

FeSe, the simplest among the known iron-based superconductors, has been the focal point of intensive study even after more than a decade of scientific research. It is believed that the simplicity of the structure may shed more light on how to clarify the mechanism of superconductivity in these compounds. The most interesting issue here is connected with recovering of the structure of the superconducting (SC) gap(s), which may be the key to revealing the pairing interaction mechanism and understanding the nature of the SC state. The most direct techniques to get information about the SC gap are spectral methods, such as angle-resolved photoemission spectroscopy (ARPES), scanning tunneling spectroscopy (STS) and point-contact (PC) Andreev reflection (PCAR) spectroscopy.

The most recent investigations of FeSe by using the above methods [1-11] allowed to make a general conclusion that FeSe is a multiband superconductor, where the SC gaps reveal anisotropic properties. Hashimoto *et al.* [1], utilizing polarization-dependent laser-excited ARPES, reported that the SC gap had a twofold in-plane anisotropy at the zone-centered hole Fermi surface. They have found considerable difference between the multi- and single-domain FeSe samples. The SC gap drops steeply to zero in a narrow angle range for the single-domain samples, evidencing for nascent node, whereas, in contrast, the multi-domain samples show finite gaps at any angle. Kushnirenko *et al.* [2] also found anisotropic SC gaps on hole- and electronlike Fermi surfaces in all momentum directions by ARPES. The in-plane anisotropy of the SC gap was explained by both nematicity-induced pairing anisotropy and orbital-selective pairing, while the $k_z$ anisotropy remains uncertain at the moment. Rhodes *et al.* [3], using high-resolution ARPES, have found that on both hole and electron pockets of the Fermi surface, the magnitude of the gap follows the distribution of $d_{yz}$ orbital weight, which, in their opinion, confirms the picture of spin fluctuation mediated superconductivity in FeSe.

One of the first STS studies of FeSe crystalline films by Song *et al.* [4] reports a gap function with nodal lines as it stems from V-shaped zero-bias minimum between the gap peaks of the tunneling spectra (*dI/dV)*. The SC gap of 2.2 meV was measured to be half as the distance between the peak positions. Onwards, Kasahara *et al.* [5] showed similar features of the tunneling spectra with evident shoulders of the multigap structure, indicating the presence of at least two superconducting gaps with $\Delta \approx 2.5$ and 3.5 meV. Later, Jiao *et al*. [6] reported multigap superconductivity based on their STS measurements on FeSe single crystals, where the isotropic *s*-wave gap is much smaller than the anisotropic *s*-wave gap of the type $\Delta^0_{es}(1 + \alpha \cos 4\Theta)$. It was shown, that SC gap also remains nodeless on twin boundaries. Sprau *et al.* [7], using sub-kelvin Bogoliubov quasiparticle interference imaging, found indications that both gaps are extremely anisotropic, yet nodeless with gap maxima oriented orthogonally in the momentum space. Such a complex gap configuration reveals the existence of the orbital-selective Cooper pairing.

Early PCAR measurements were performed on break junctions in polycrystalline FeSe samples [8]. The *dI/dV* spectra revealed two sets of subharmonic gap structures due to multiple Andreev reflection. This was taken as a proof of the presence of two nodeless SC gaps $\Delta_L = 2.75 \pm 0.3$ meV and $\Delta_S = 0.8 \pm 0.2$ meV. Our PCAR studies of FeSe single crystals are published in series of papers [9-11]. We extracted two gaps from the measured *dV/dI* spectra [9] with gap values similar to those observed in STS experiments [4, 5]. Along with this, in some PC's we observed the increase in the critical temperature by more than two times [10]. A more detailed study of the SC gap behavior in FeSe was presented in [11] by using the method of "soft" PCAR spectroscopy. Analyzing the $dV/dI(V)$ spectra for 25 (PCs) we obtained the average gap values $\Delta_L = 1.8 \pm 0.4$ meV and $\Delta_S = 1.0 \pm 0.2$ meV giving the reduced values of the superconducting coupling strength $2\Delta_L/k_BT_c = 4.2 \pm 0.9$ and $2\Delta_S/k_BT_c = 2.3 \pm 0.5$ for the large ($L$) and small ($S$) gap, respectively. The temperature dependencies of both gaps revealed standard BCS like behavior. Additionally, a small gap contribution was found to be within tens of percent, decreasing with both temperature and magnetic field. The lowest temperature in the above-mentioned study was only 3K. At the same time, we found in [11] that the *dV/dI* spectra at 3K may be equally well described by theoretic curves using an anisotropic single-gap model and a model with two superconducting gaps. It is known that the temperature is responsible for the resolution in PCAR spectroscopy[1]. Therefore, measurements at lower temperatures are desirable to get more correct information about the absolute value and anisotropy of the SC gaps. Thus, in the present study PCAR measurements were carried out at lower temperatures down to 0.5K.

## Method

The "soft" PCs were made by placing a tiny drop of a silver paste between the FeSe sample and a 0.1 mm copper wire. The same technique was used in [11]. Four of these contacts made to one crystal are shown in Fig.1inset. A standard 20-lead chip was used as the sample substrate, enabling simultaneous measurements on several "soft" PCs during one cooling down period. Fig. 1 demonstrates four wires each soldered with two leads on a chip in order to realize "pseudo" four probe measuring configuration since the sample was also electrically coupled with two separate leads. The chip itself was clamped to a cold plate of a $^3$He refrigerator at ambient conditions. After cooling down the temperature of the chip was measured and controlled with a multi-channel controller, which regulates the temperature in the 0.5-20 K range.

The FeSe samples where grown by flux technique method. The description of the crystal growth can be found in [12]. In our experiments we measured PCAR spectra (differential resistance *dV/dI(V))* through a "soft" PC by a standard modulation method. The schematic picture of measuring setup is shown in Figure 4.3 of Ref. [13].

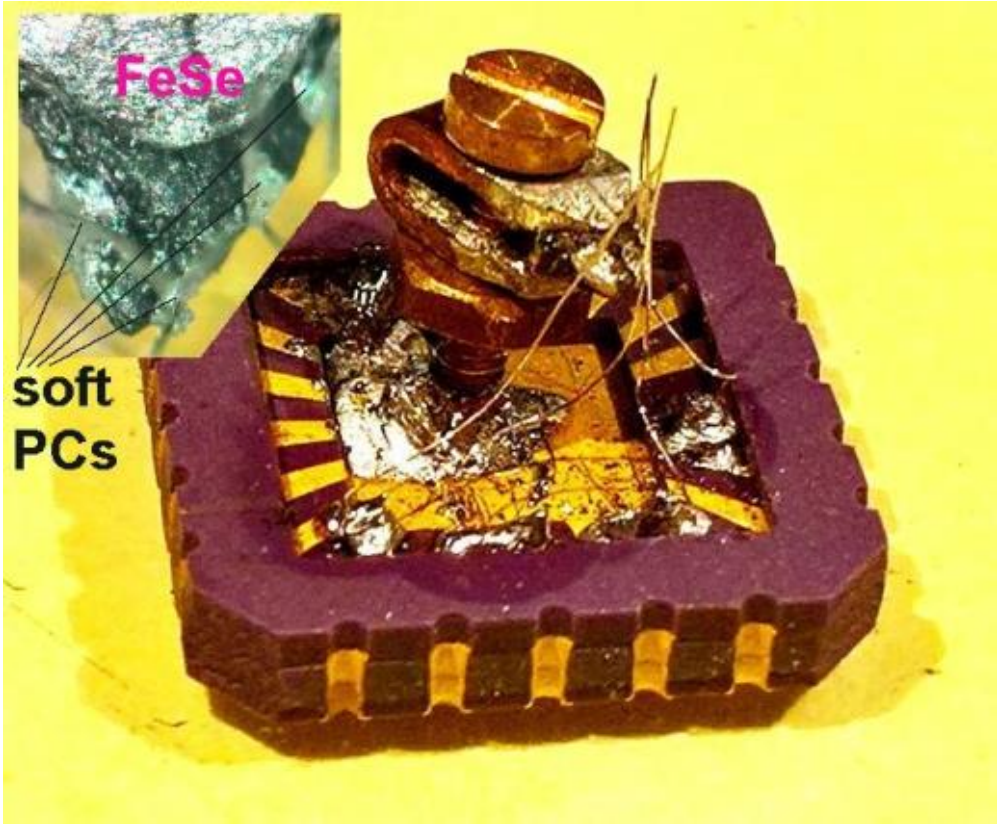

**Fig. 1**.
FeSe sample clamped within the copper clip by means of the brass screw and the nut. To create good thermal contact the screw itself was soldered to the base plate of the 20-lead chip and coupled electrically with two such leads. Inset (left top corner): Enlarged image of the FeSe sample with four "soft" contacts attached.

## Results and discussion

Figure 2 shows temperature a), c) magnetic field b), d) variation of the *dV/dI* spectra for two representative "soft" PCs. PC spectra measured at the lowest temperatures show a double-minimum structure characteristic for the Andreev reflection. This structure is suppressed as the temperature increases, as it is seen in Figure 2a) and c). At first, it transforms into a single minimum which vanishes approaching the mark of 12 K similarly to our previous study [11]. This $T_c$ is a few degrees higher than the bulk critical temperature $T_c$=9.4K [12]. Higher values of the local $T_c$'s were observed in our previous paper [10] where we explained different surface and bulk properties on account of interfacial effects. Figures 2 b) and d) show the magnetic field dependencies of the PC spectra from a) and c) measured at the lowest possible temperature T = 0.5 K. Both sets reveal similar field dependencies. In magnetic fields up to 8 T the filling of the gap structure is evident and surprisingly, the suppression of the double-minimum intensity is not followed with the moving of the gap minima to lower energies, as it is usually observed in classical BCS superconductors. Unfortunately, the largest magnetic field B = 8 T is much lower than the upper critical magnetic field of the studied samples, which is above 20T at T=0.5K[14]. Thus, we cannot follow the full suppression of the superconducting gap(s). Our PCAR spectra, shown in Figure 2, reveal additional spectral features in the form of symmetric side maxima above the gap structure at around 3 mV. Fig. 2a), b) clearly depicts these maxima, while in Fig. 2c), d) they look just like shoulders. The presence of side maxima above the gap structure of the *dV/dI* spectra is a typical signature of the suppression of superconductivity with high current density and/or temperature increase due to the transition to the thermal regime with a bias rise of a non ballistic, likely diffusive PC [15, 16].

[1] The energy resolution for point-contact spectroscopy is determined by two factors - the temperature $T$ and the modulating signal intensity $V_{mod}$, if a synchronous detection of the first derivative of the *I-V* curve is used. As a result, the resolution of the first derivative *dV/dI* calculated by relation $\Delta\varepsilon=\sqrt{(3.53k_BT)^2+(2.45V_{mod})^2}$ [13] is between 160-170 μeV at 0.5K by using typical $V_{mod}$=20-30 μeV. Thus, even at sub-kelvin temperature the main resolution is determined by the temperature.

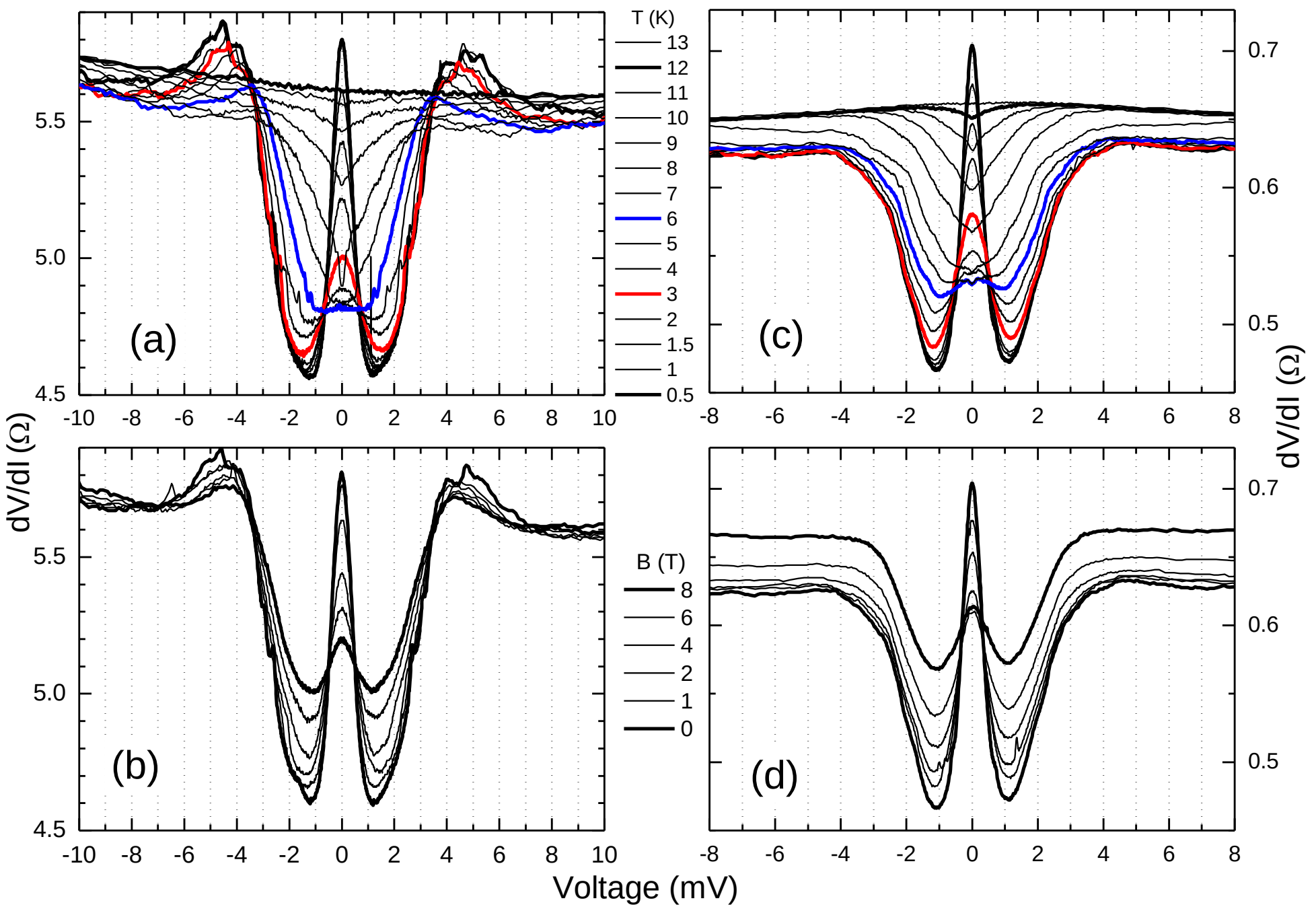

**Fig. 2**. Temperature and magnetic field variation of the differential resistance *dV/dI* for two “soft” PCs measured in the range of the lowest temperature of 0.5K to 13K and in the magnetic field up to 8T at 0.5K.

The measured *dV/dI* spectra have been fitted to the Blonder-Tinkham-Klapwijk (BTK) equation [11,17], which describes the Andreev reflection of quasiparticles at the ballistic PC between a normal metal and a superconductor taking into account the superconducting gap Δ, the PC barrier strength Z and the spectral smearing parameter Γ. We focused on the spectra which revealed the lowest presence of the disturbing side maxima (from Fig. 2c) and d)). We also fitted our data with three modifications of the BTK model, taking into account an anisotropic s-wave gap, and the weighted sums of two isotropic or anisotropic gaps. While fitting data, we considered the so-called scaling parameter S which is included to fit the intensity of the calculated and experimental curves, that is $S=(dV/dI)_{\mathrm{exp}}/(dV/dI)_{\mathrm{theor}}$. For instance, $S = 1$ means that the calculated curve also fits the measured *dV/dI* in absolute values[2], in other words, it fits its intensity.

In the case of fitting to the anisotropic one gap model, the energy gap and the spectral smearing parameter were defined with the anisotropy functions $\Delta=\Delta^0(1 +\alpha \cos 4\Theta)$ [6], $\Gamma= \Gamma^0(1 +\alpha \cos 4\Theta)$, where $\alpha$ is the anisotropy parameter. The fitting parameters are shown in the legends for Figure 3b. The fitting curve is shown as a solid red curve. The quality of fit is sufficient, however, the scaling parameter S >>1.

[2] Some details of the fitting procedure and the meaning of the fitting parameters are described in Appendix of Ref. [18].

When we applied a weighted sum of two anisotropic gap contributions, the number of fitting parameters redoubled due to the contribution of the second gap. We reduced the number of fitting parameters, supposing that Z was the same for both gaps because this parameter reflects primarily the physical character of the interface, rather than details of the electronic structure as discussed in [16]. Also, the same anisotropy $\alpha$ has been predicted for both gaps. While fitting data we used the weight parameter *w*, which characterizes the weight of the small gap and (1-*w*) the large gap contribution. The fitting result is shown in Figure 3a. The scaling parameter S reveals the value close to 1.

In the case of the isotropic two-gap model we simply put $\alpha$ =0, as described in our previous paper [11]. The fitting curve plotted in Figure 3 c) as a red curve describes our experimental data similarly to the anisotropic one gap fit (see Figure 3c), but then again with the scaling parameter S>>1.

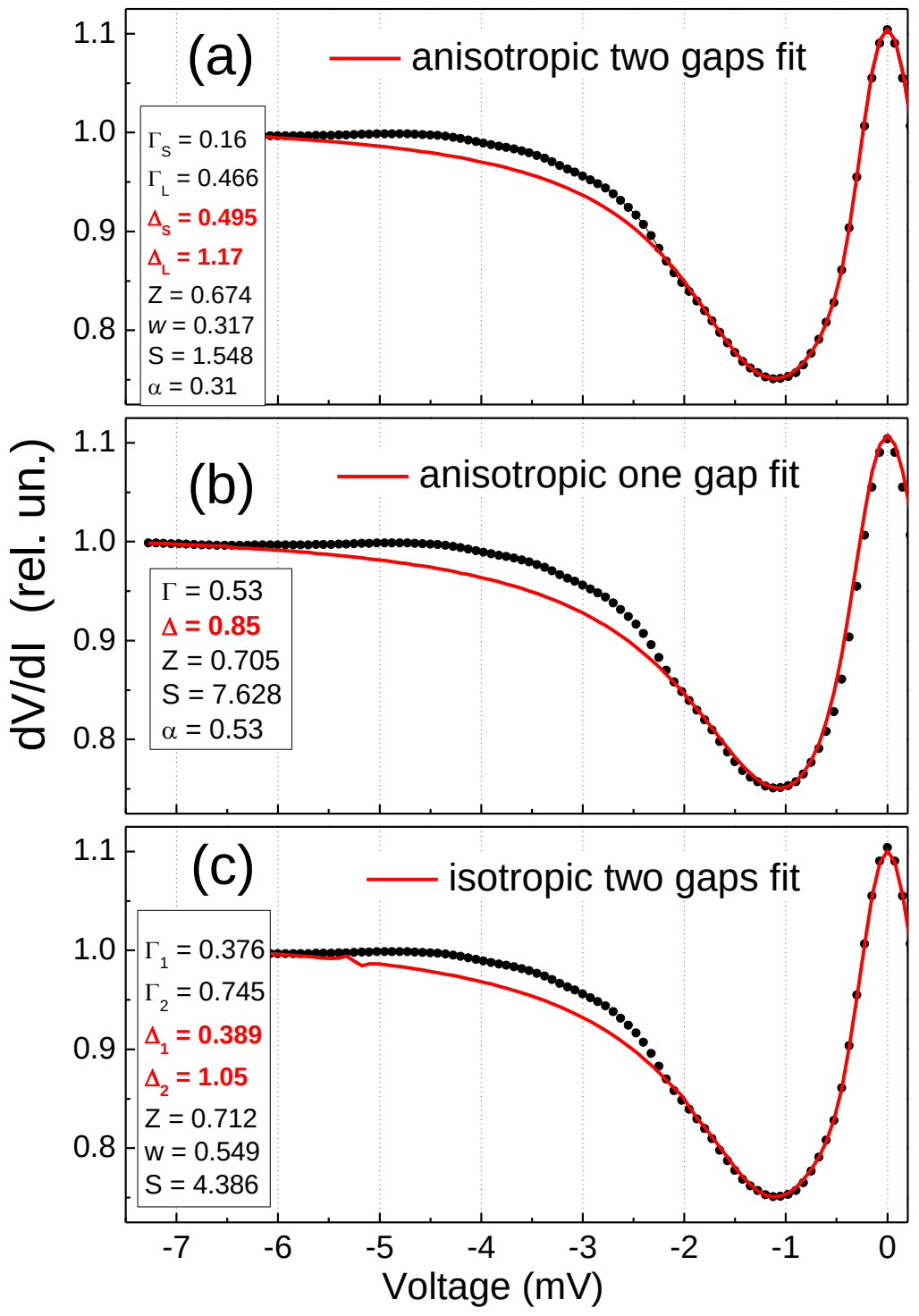

**Fig. 3.** Comparison of different fit models at temperature 0.5K. The experimental data are shown as symbols and fits are shown as a solid (red) line. Legends in each panel show the fitting parameters.

Fig. 3 offers the possibility to compare fitting results of the used one- and two-gap models at T = 0.5 K. All models may help to describe sufficiently accurate experimental curves. The dominant gap value in all curve fittings is close to the position of the *dV/dI* spectral minimum, which is positioned just above 1 mV in the measured spectra. However, only the anisotropic two gap fit gives the scaling parameter S near to 1. It is important to notice that in this fit we obtained the smallest value for the spectral smearing parameters $\Gamma^0{}_S$ and $\Gamma^0{}_L$. Therefore, the anisotropic two-gap model may in our opinion fit our spectra more accurate. Discrepancies may be visible only between 3 and 6 mV, where the non-Andreev-reflection like side maxima disturb the spectra. These maxima are strongly sensitive to the measuring temperature and the applied magnetic field. This is visible in Figure 2 where the maxima are shifting to lower energies (into the gap) at increased temperatures and magnetic fields. The presence of this disturbing effect may strongly affect the study of the gap features of the *dI/dV* spectra at higher temperatures. Therefore, we were able to perform an adequate fit and to get temperature dependence of the fitting parameters only up to 6 K.

Figure 4 shows all of our fitting results. Figure 4a) plots the temperature dependencies of the superconducting gaps with discrete symbols. The gray part of it shows areas of the possible extrapolations of the data with standard BCS curves. Possible values of the critical temperature Tc are between the bulk value 9.4K and 12K when the SC minimum disappears. Note, that a non-Andreev V-type shape of zero-bias minimum above 9K (see Fig. 2c) most likely testifies to transition to the gapless superconducting fluctuation region. Therefore, BCS extrapolation of the gaps to the bulk critical temperature is more accurate. It should also be mentioned that the anisotropy parameter $\alpha$ has a value 0.31 at low temperature, which is close to 0.34 reported by [6].

The fitting parameters of the spectral weight *w*, spectral smearing $\Gamma^0_S$ and $\Gamma^0_L$ and gap anisotropy $\alpha$ reveal evident temperature dependencies above 3 K (see Fig.4). These dependencies are probably connected with the strong temperature dependence of the side maxima (positioned at 3 mV at T 0.5K), which are moving into the gap at increased temperatures disabling correct fitting of the spectral features. The weak temperature dependence of the scaling parameter S (between 1.6 and 1), shown in Figure 4b) as red symbols and an almost temperature independent value of the barrier strength (black symbols in Figure 4b)) is a confirmation of the validity of our fitting model with the weighted sum of two Andreev reflection contributions with anisotropic energy gaps. The magnetic field dependence of the PC spectra, shown in Figure 2 b) and d) cannot be observed in a single band system, thus, it represents another proof of the multiband superconductivity in FeSe samples.

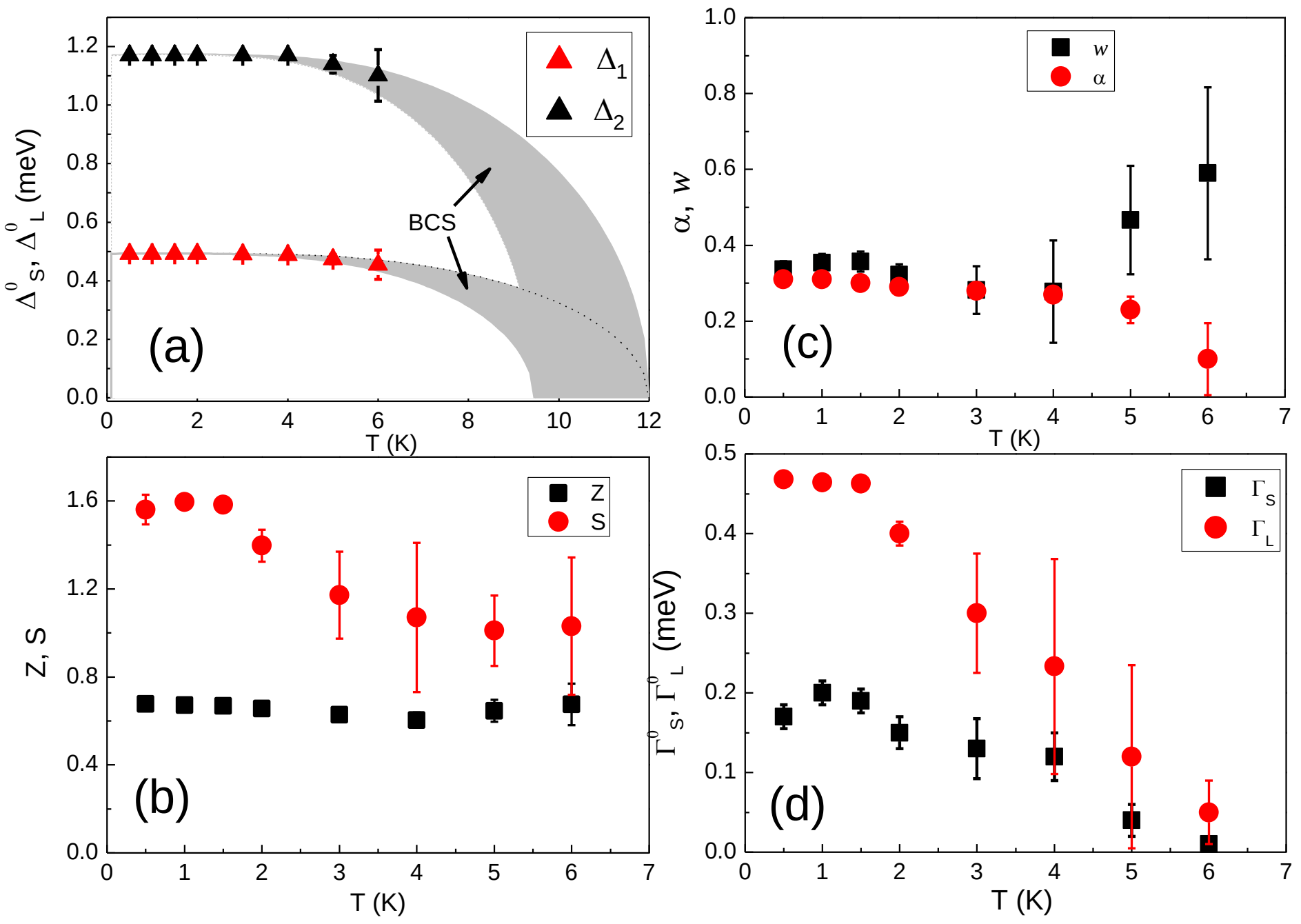

**Fig.4**. Calculated temperature dependence of the fitting parameters: SC gaps $\Delta^0_{L,S}$, broadening parameter $\Gamma^0_{L,S}$, barrier Z and scaling parameter S, weight factor *w* and anisotropic parameter $\alpha$. Shadow area in panel (a) is BCS-like dependence corresponding to the critical temperature between the bulk value of 9.4K in FeSe and 12K, the point at which the zero-bias minimum of *dV/dI* disappears.

With regard to the measurements in a magnetic field, the available field strength was too weak to suppress superconductivity and the field effect was mainly observed in the broadening of the curves (see Fig. 2b, d). As it was shown in [11], at least a two times larger field is needed to practically suppress the double-minimum structure. Anyway, some noteworthy results are as follows: similar to what was found in Ref. [11], the SC gap value is robust[3] with respect to the magnetic field strength up to 8T although contribution *w* of the small gap vanishes, which is consistent with the observations made in our previous paper [11]. Besides, anisotropy parameter $\alpha$ turned out to be field independent.

## Conclusion

We have performed PCAR measurement in FeSe single crystals in the sub-kelvin temperature range. It allowed us to perform high resolution Andreev-reflection measurements in this compound. Analyzing our PCAR *dV/dI* data we have shown that two anisotropic superconducting gaps are responsible for the superconducting properties of FeSe. The temperature dependencies of the superconducting gaps determined by *dV/dI* curve fitting strongly support the anisotropic two-band scenario of superconductivity in this compound .

## Acknowledgments

The studies were conducted as part of a joint project between National Academy of Sciences of Ukraine and Slovak Academy of Sciences. DLB and NVG would like to thank the Institute of Experimental Physics (Košice) for their hospitality, O.E. Kvitnitskaya for useful discussions and D. Chareev for providing FeSe samples. The supports of Slovak APVV-16-0372 and VEGA 2/0149/16 projects are acknowledged.

## Literature:

1. T. Hashimoto, Y. Ota, H. Q. Yamamoto, Y. Suzuki, T. Shimojima, S. Watanabe, C. Chen, S. Kasahara, Y. Matsuda, T. Shibauchi, K. Okazaki, and S. Shin, *Superconducting gap anisotropy sensitive to nematic domains in* FeSe, Nat. Commun.**9**, 282 (2018).

2. Y. S. Kushnirenko, A. V. Fedorov, E. Haubold, S. Thirupathaiah, T. Wolf, S.Aswartham, I. Morozov, T. K. Kim, B. Büchner, and S. V. Borisenko, *Three-dimensional superconducting gap in* FeSe *from angle-resolved photoemission spectroscopy*, Phys. Rev. B **97**, 180501 (2018).

3. Luke C. Rhodes, Matthew D. Watson, Amir A. Haghighirad, Daniil V. Evtushinsky, Matthias Eschrig, and Timur K. Kim, *Scaling of the superconducting gap with orbital character in* FeSe, Phys Rev B, **98**, 180503(R) (2018).

4. C.-L. Song, Y.-L. Wang, P. Cheng, Y.-P. Jiang, W. Li, T. Zhang, Z. Li, K. He, L. Wang, J.-F. Jia, H.-H. Hung, C. Wu, X. Ma, X. Chen, and Q.-K. Xue, *Direct observation of nodes and twofold symmetry in* FeSe *superconductor*, Science **332**, 1410(2011).

5. S. Kasahara, T. Watashige, T. Hanaguri, Y. Kohsaka, T. Yamashita, Y. Shimoyama, Y. Mizukami, R. Endo, H. Ikeda, K. Aoyama, T. Terashima, S. Uji, T. Wolf, H. v. Löhneysen, T. Shibauchi, and Y. Matsuda, *Field-induced superconducting phase of* FeSe *in the BCS-BEC crossover*, Proc. Natl. Acad. Sci. USA **111**, 16309 (2014).

6. L. Jiao, C.-L. Huang, S. Rößler, C. Koz, U. K. Rößler, U. Schwarz, and S. Wirth, *Superconducting gap structure of* FeSe, Sci. Rep. **7**, 44024 (2017).

7. P. O. Sprau, A. Kostin, A. Kreisel, A. E. Böhmer, V. Taufour, P. C. Canfield, S. Mukherjee, P. J. Hirschfeld, B. M. Andersen, J. C. S. Davis, *Discovery of orbital-selective Cooper pairing in* FeSe, Science **357**, 75–80 (2017).

---

[3] Similar behavior of delta in magnetic field was observed for $TmNi_2B_2C$ [18]. One of the possible explanations could be that if the vortices are pinned near the contact, e.g. on the edges, then the gap remains almost unperturbed inside the contact.

8. Ya. G. Ponomarev, S. A. Kuzmichev, M. G. Mikheev, M. V. Sudakova, S. N. Tches-nokov, T. E. Shanygina, O. S. Volkova, A. N. Vasiliev, and Th. Wolf, *Andreev spectroscopy of* FeSe: *Evidence for two gap superconductivity*, J. Exp. Theor. Phys. **113**, 459 (2011).

9. Yu. G. Naidyuk, N. V. Gamayunova, O. E. Kvitnitskaya, G. Fuchs, D. A. Chareev, and A. N. Vasiliev, *Analysis of nonlinear conductivity of point contacts on the base of* FeSe *in the normal and superconducting state*, Fiz. Nizk. Temp. **42**, 42 (2016) [Sov. J. Low Temp. Phys. **42**, 31 (2016)].

10. Yu. G. Naidyuk, G. Fuchs, D. A. Chareev, and A. N. Vasiliev, *Doubling of the critical temperature of* FeSe *observed in point contacts*, Phys. Rev. B **93**, 144515 (2016).

11. Yu. G. Naidyuk, O. E. Kvitnitskaya, N. V. Gamayunova, D. L. Bashlakov, L. V. Tyutrina, G. Fuchs, R. Hühne, D. A. Chareev, and A. N. Vasiliev, *Superconducting gaps in* FeSe *studied by soft point-contact Andreev reflection spectroscopy*, Phys. Rev. B **96**, 094517 (2017).

12. D. Chareev E. Osadchii, T. Kuzmicheva, J.-Y. Lin, S. Kuzmichev, O. Volkova, and A. Vasiliev, *Single crystal growth and characterization of tetragonal* $FeSe_{1-x}$ *superconductors*, CrystEngComm **15**, 1989 (2013).

13. Yu. G. Naidyuk and I. K. Yanson, *Point-Contact Spectroscopy*, Springer Series in Solid-State Sciences (Springer Science+Business Media, Inc), vol. 145, 2005.

14. S. I. Vedeneev, B. A. Piot, D. K. Maude, and A. V. Sadakov, *Temperature dependence of the upper critical field of FeSe single crystals*, Phys. Rev. B **87**, 134512 (2013).

15. Goutam Sheet, S. Mukhopadhyay, and P. Raychaudhuri, *Role of critical current on the point-contact Andreev reflection spectra between a normal metal and a superconductor*, Phys. Rev. B **69**, 134507 (2004).

16. Yu. G. Naidyuk, K. Gloos, *Anatomy of point-contact Andreev reflection spectroscopy from the experimental point of view*, Fiz. Nizk. Temp. **44**, 343 (2018) [Low Temp. Phys. **44**, 257 (2018)].

17. G.E. Blonder, M. Tinkham, and T.M. Klapwjik, *Transition from metallic to tunneling regimes in superconducting microconstrictions: Excess current, charge imbalance, and supercurrent conversion*, Phys. Rev. B **25**, 4515 (1982).

18. Yu. G. Naidyuk, O.E. Kvitnitskaya, L.V. Tiutrina, I.K. Yanson, G. Behr, G. Fuchs, S.-L. Drechsler, K. Nenkov, L. Schultz, *Peculiarities of the superconducting gaps and the electron-boson interaction in* $TmNi_2B_2C$ *as seen by point-contact spectroscopy*, Phys. Rev. B, **84**, 094516 (2011).